\documentclass[12pt,a4paper,english,amssymb,nofootinbib,superscriptaddress]{revtex4-2}
\usepackage{xcolor}
\usepackage{lmodern}
\usepackage{lmodern}

\usepackage[T1]{fontenc}
\usepackage[latin9]{inputenc}
\usepackage{stmaryrd}
\usepackage{babel}
\usepackage{fancybox}
\usepackage{calc}
\usepackage{amsmath}
\usepackage{amssymb}
\usepackage{esint}
\usepackage[pdfusetitle,
 bookmarks=true,bookmarksnumbered=false,bookmarksopen=false,
 breaklinks=false,pdfborder={0 0 1},backref=false,colorlinks=false]
 {hyperref}

\makeatletter

\usepackage{babel}

\def\be{\begin{equation}}
\def\ee{\end{equation}}

\@ifundefined{textcolor}{}{%
	\definecolor{BLACK}{gray}{0}
	\definecolor{WHITE}{gray}{1}
	\definecolor{RED}{rgb}{1,0,0}
	\definecolor{GREEN}{rgb}{0,1,0}
	\definecolor{BLUE}{rgb}{0,0,1}
	\definecolor{CYAN}{cmyk}{1,0,0,0}
	\definecolor{MAGENTA}{cmyk}{0,1,0,0}
	\definecolor{YELLOW}{cmyk}{0,0,1,0}
}

\usepackage{latexsym}\usepackage{bm}

\makeatother

\begin{document}

\title{Physical Interpretations of Integration Constants and Large Gauge Effects in Flat and AdS Spacetimes}
\author{Leyla Ogurol}
\email{leyla.ogurol@metu.edu.tr}

\affiliation{Department of Physics,~\\
 Middle East Technical University, 06800, Ankara, Turkey}
\author{Bayram Tekin\thanks{Corresponding author} }
\email{btekin@metu.edu.tr}

\affiliation{Department of Physics,~\\
 Middle East Technical University, 06800, Ankara, Turkey}
\date{\today}

\begin{abstract}

\noindent As in other partial differential equations, one ends up with some
arbitrary constants or arbitrary functions when one integrates
Einstein's equations, or more generally field equations of any other
gravity. Interpretation of these arbitrary constants and functions
as some physical quantities that can in principle be measured is a non-trivial matter. Concentrating on the case of constants, one usually identifies them as conserved mass, momentum, angular momentum, center of mass, or some other hairs of the solution. This can be done via the Arnowitt-Deser-Misner (ADM)-type construction based on pure geometry, and the solution is typically a black hole. Hence, one talks about the black hole mass and angular momentum, etc. Here we show that there are several misunderstandings: First of all, the physical interpretation of the constants of a given geometry depends not only on pure geometry, i.e. the metric, but also on the theory under consideration. This becomes quite important, especially when there is a cosmological constant. Secondly, one usually assigns the maximally symmetric spacetime, say the flat or the (anti)-de Sitter spacetime, to have zero mass and angular momentum and linear momentum. This declares the maximally symmetric spacetime to be the vacuum of the theory, but such an assignment depends on the coordinates in the ADM-type constructions and their extensions: in fact, one can introduce large gauge transformations (new coordinates) which
map, say, the flat spacetime to flat spacetime but the resultant flat spacetime can have a nontrivial mass and angular momentum, if the new coordinates are such that the metric components do not decay properly. These
issues, which are often overlooked, will be examined in detail, and a resolution, via the use of a divergence-free rank $(0,4)$-tensor, will be shown for the case of anti-de Sitter spacetimes.
\end{abstract}
\maketitle

\begin{center}
To appear in a special issue of \textit{Nuclear Physics B} :\\
\textit{``Clarifying Common Misconceptions in High Energy Physics and Cosmology''},\\
Eoin O Colgain and Yasaman Farzan, eds.
\end{center}

\clearpage
\tableofcontents
\clearpage

\section{Introduction}

The history and development of physics can also be read from the point of view of our understanding of the concept of energy and its various forms or appearances/representations. We have learned that this conserved quantity in a closed system can take different forms through interactions. The standard model fields, all quantum fields, are representations of energy, momentum, etc. From this vantage point, an electron is an electronic field, a fermionic form of energy-momentum, spin, and other conserved quantities, which appears in the standard model Hamiltonian and interacts with other forms of energy, momentum, etc., as dictated by the Hamiltonian. Of course, the full quantum theory allows infinitely many interactions as the time-evolution operator involves not just the Hamiltonian, but the exponential of the Hamiltonian, which in perturbative expansion produces all sorts of higher-order interactions. All this is well understood and rigorously defined through Noether's theorem \cite{1}, both at the classical and quantum level (where one needs renormalization and regularization). Rigid symmetries of spacetime (these are isometries of Minkowski spacetime) lead to conservation laws pertaining to energy and momentum, angular momentum, and typically, we do not expect these symmetries and conservation laws to be broken in quantum theory. But once gravity is introduced through geometry as in the case of Einstein's General Relativity (GR) and the spacetime becomes a physical entity in the sense that there is a gravitational field $g_{\mu \nu}(x)$ that also can carry energy, momentum, spin, etc., Noether's theorem, connecting symmetries and conserved charges immediately gets modified in a significant way. As we said elsewhere \cite{Adami}: "The first casualty of gravity" is Noether's theorem, as far as the rigid symmetries are conserved. A generic solution of Einstein's equations need not have any isometries, hence, we cannot really talk about energy conservation in a spacetime that does not have a time-like Killing vector. 

The problem of defining conservation laws when the metric field is also a representation (or a form) of energy-momentum, etc., is challenging because of the way gravity is modeled to arise as an effect of curvature on a Riemannian manifold. In a Locally Inertial Frame (LIF), we can make the metric trivial at a point $p$ in the sense that we have
\begin{equation}
g_{\hat{\mu} \hat{\nu}}\vert_p= \eta_{\hat{\mu} \hat{\nu}} \hskip 2 cm  \partial_{\hat{\sigma}}g_{\hat{\mu} \hat{\nu}}\vert_p=0, \label{LIF}
\end{equation}
which is a beautiful way to eliminate local gravity and make it in some sense a tidal effect of which the measurable consequences appear in the curvature since coordinates do not allow us to set the second derivative of the metric to vanish: $\partial_{\hat{\rho}}\partial_{\hat{\sigma}}g_{\hat{\mu} \hat{\nu}}\vert_p \ne 0$. So we must be able to answer the following question: What is the energy-momentum tensor of the gravitational field $g_{\mu \nu}(x)$ under the condition that this field is trivial locally as given in (\ref{LIF})? It is clear that no such meaningful (tensorial) expression can exist, because, as a spin-2 bosonic field, one would expect the energy-momentum tensor of the gravitational field should be locally like 
\begin{equation}
   T_{\mu \nu} \sim \left(\partial g\partial g \right)_{\mu \nu} +...,
\end{equation}
which is zero in the LIF system, and hence, as a tensor, it must be zero in all coordinates. One could try to circumvent this impasse and write it as $T_{\mu \nu} \sim \left( \eta \partial\partial g\right)_{\mu \nu} +...$, but it has a bimetric form and cannot be completed to a tensor. All this is known and has been discussed many times in the past 110 years of the history of GR \cite{metric}. This, in some sense, is a highly non-trivial situation: When we say some field, or some particle {\it exists}, what we mean is that we can associate a non-zero, local $T_{\mu \nu}(x)$ to the free version of this field. For the case of the gravitational field, this is apparently not the case: There does not exist a bona fide tensor which represents the energy-momentum tensor of the gravitational field.\footnote{As a side remark, let us note that the problem is so acute that one cannot even write a mass term for the gravitation or a gravitational wave, as mass of the particle or field, being a form of local energy, would violate diffeomorphism invariance. The suggested ways out of this dilemma involve more than one field in four dimensions and higher curvature theories in 2+1 dimensions.} 
The resolution is the following: We have to use some approximation, either quasi-local expressions (that is, integrals over a finite region of spacetime) or global expressions that involve integrals over spacetime. If we read Einstein's equation 
\begin{equation}
 R_{\mu \nu} - \frac{1}{2}g_{\mu \nu}R + \Lambda g_{\mu \nu} = \frac{8 \pi G_N}{c^4} T_{\mu \nu},   \label{Einstein_eq} 
\end{equation}
in this light, we have the following interesting phenomenon:  Let us start with a large amount of self-gravitating gas out there in space, say described by a perfect fluid with an energy-momentum tensor of the form 
\begin{equation}
T_{\mu \nu}= \left( \rho + \frac{P}{c^2} \right) u_\mu u_\nu +  P g_{\mu \nu},   \label{Einstein_eq}
\end{equation}
and a given equation of state $P =P(\rho)$. So we have a compactly supported source and $T_{\mu \nu}$ is a local tensor field that cannot be made to vanish by choosing coordinates.  We understand this "initial" form of existence, or energy-momentum, in a fairly reasonable manner. But then we know that self-gravitation of this initial gas state will lead to the continued collapse, and no thermal equilibrium is possible since increasing the pressure will lead to a more rapid collapse due to the dual role played by pressure in gravity. If the total mass of the gas is sufficiently large, then non-thermal (quantum mechanical) pressures like the electron degeneracy or neutron degeneracy pressure also do not provide a halt to the collapse, and the system collapses to a black hole. We expect this black hole to be a Kerr black hole if there is a finite angular momentum to begin with. Kerr black hole is a {\it vacuum} solution to Einstein's equations. Therefore, the initial localized energy-momentum, angular momentum, etc., as represented or carried by the gas, at the end turns into the form represented by the metric $g_{\mu \nu}$ of spacetime. So apparently, Einstein's equation  (\ref{Einstein_eq}) should not just be seen as an equation where gravity is sourced by matter's energy-momentum tensor, but the left-hand side of the equation eats the right-hand side and turns the initial local energy-momentum tensor into a highly non-local one. In another interpretation, at the classical level, the left-hand side is the future of the right-hand side.  Of course, when quantum physics is introduced, the formed black hole will emit Hawking radiation, and the highly non-local energy in the form of the black hole (or in the form of a non-trivial curvature distribution) will turn into radiation and matter, which is again locally non-trivial. 

All these considerations and many more suggest that when the local energy-momentum turns into the form represented by the metric tensor, one must necessarily identify some large portions of spacetime, or all of it, to define the conserved quantities. Of course, as the arguments above suggest, it would be best if the second derivatives of the metric tensor are involved in these expressions, not the metric tensor and its first derivative; the latter two can be chosen to be trivial at a point. The correct collection of the second derivatives of the metric is in the form of the Riemann curvature tensor $R^\mu\,_{\nu \sigma \rho}$ since this is a proper tensor; and it is the necessary and sufficient tensor to describe the curvature of spacetime in dimensions greater than 3. This can be seen in various ways, but let us just note the following: In the Riemann-normal coordinates (these are the coordinates built in a small neighborhood around a point $p$ by using the non-crossing geodesics that emanate from that point in all directions), the metric in Taylor series expansion around the point reads (see for example \cite{Brewin})
\begin{equation}
g_{\mu \nu}(x) = g_{\mu \nu}(p) -\frac{1}{3}x^\sigma x^\rho R_{\mu \sigma \nu\rho}(p) - \frac{1}{6}x^\sigma x^\rho  x^\lambda \nabla_\sigma R_{\mu \rho \nu\lambda}(p)+ {\mathcal{O}}(x^4). \label{normal1}
\end{equation}
So, this expression gives us the non-trivial local content of the metric field: the zeroth-order term is a constant, while the first-order term is nonexistent, as expected, while the second-order and higher-order terms are organized in derivatives and the powers of the Riemann tensor. The fact that the Ricci tensor does not appear in (\ref{normal1}) is important: Einstein's theory must allow for nonzero curvature in the regions where there is no matter ($T_{\mu \nu}=0$), otherwise matter cannot curve the spacetime out side of it. But once this is allowed, the gravitational field itself can also curve spacetime, i.e., the theory is non-linear, and hence the energy-momentum tensor of the gravitational field is non-localizable. So one barters non-trivial gravity outside a matter source with non-linearity. Einstein tensor appears in the $\det(g_{\mu \nu})$ as can be seen from (\ref{normal1}) only inside the matter. Equation (\ref{normal1}) gives us hope that we can consider the metric field just like a regular field, but it also gives us the warning that if we want to calculate the conserved charge content of the metric field, we will have to consider the total universe for which this truncated Taylor series is not sufficient. In fact, as the $x$ coordinate becomes large, the geodesics will cross each other, and we will not be able to use them as a coordinate system. However, the hint we get from that equation will lead to an expression of conserved charges in terms of the curvature for asymptotically constant curvature space-times, as we shall see later. 

Our current understanding, when it comes to conserved quantities in gravity theories, is briefly as follows. We take space-time as globally hyperbolic
(as this is the case suitable for the Cauchy initial value problem and as the time-like Killing vector can exist), $\mathcal{M} = \mathbb{R}\times \Sigma$, where by equality we mean diffeomoprhic equivalence; and the boundary of the spatial part $\Sigma$ is nonempty. [In fact, to allow for non-trivial configurations such as black holes, we must allow the asymptotic region at spatial infinity, that is, the asymptotic boundary of $\Sigma$, to be composed of more than one asymptotic end.] Then, heuristically from far away, each spatial hypersurface of the universe basically looks like a particle, albeit an extended one, with energy-momentum, etc. Defining conserved energy and momentum that do not depend on the coordinates is one problem; having a consistent, i.e., a bounded-from-below result, is another problem. Especially the positive mass and energy issue of the ADM \cite{adm} mass, which not only describes a physical, isolated system, from a distance, but also a geometric invariant assigned to an asymptotically flat manifold, turned out to be a highly nontrivial problem since the early 1960s and was eventually settled by Schoen and Yau \cite{Schoen:1979wk} via the techniques from minimal surface theory and the Jang equation \cite{Jang} that work for manifolds of dimension $n \le 7$. Witten \cite{Witten} gave a more physically transparent proof based on supergravity that works on all spin manifolds.

ADM mass/energy definition does not work for asymptotically constant curvature space-times and does not represent the conserved quantity in theories other than General Relativity. To remedy this, Abbott and Deser constructed the conserved charges in the cosmological Einstein theory in \cite{AD}, which was then extended to generic theories of gravity in \cite{Deser_Tekin-PRL,Deser_Tekin-PRD}. For theories with a Lagrangian of the form ${\mathcal{L}}= F(R^{\mu\nu}\,_{\sigma \rho})$ where $F$ is a smooth function of the Riemann tensor, see \cite{Senturk} and for a recent pedagogical discussion, see \cite{Bayram} and the references therein. In constant curvature backgrounds, generically, all the terms in the Lagrangian contribute to all of the conserved charges: the metric alone does not yield the conserved charges. 
Of course, the crucial point here is that conserved quantities in these theories must satisfy the first law of black hole thermodynamics, which is also a statement of energy conservation.

Let us start with the asymptotically flat and constant-curvature spacetimes. For the ADM mass, we can forget about space-time and just concentrate on the spatial part ($\Sigma$) at time $t$, which we assume is asymptotically flat. The metric outside a large ball reads $g_{ij} = \delta_{i j}+ h_{ij}$ in Cartesian coordinates. The ADM mass of this co-dimension one spatial hypersurface is given as 
\begin{eqnarray}
M_{{\rm ADM}}=\frac{1}{4\,\Omega_{n-2}\,G_N}\int_{S^{n-2}} & dS_{i}\, & \left(\partial_{j}\,h^{ij}-\partial^{i}\,h_{jj}\right),\label{adm1}
\end{eqnarray}
which is valid in the mentioned coordinates and under certain decay conditions to be discussed below. This formula also clearly assigns zero mass to the globally flat (Euclidean) metric, as expected. But, of course, as we shall see, problems arise when we change the coordinates, even without changing the geometry. This is because the splitting $g_{ij} = \delta_{i j}+ h_{ij}$ is clearly coordinate-dependent: There are two types of coordinate transformation that we can envisage: Rigid ones $x'^i = R^i\,_j x^j$, with $R^i\,_j$ being constant $SO(n-1)$- matrices that keep the background $\delta_{i j}$ intact and act on the deviation as $h'_{ij} = R^k\,_i R^m\,_j h_{k m}$. The second type are generic (albeit infinitesimal, $x^i \rightarrow x^i + \zeta^i(x)$)  diffeomorphisms that act as $\delta_\zeta h_{ij}= -\partial_i \zeta_j - \partial_j \zeta_i$. Large diffeomorphisms are the subject of this work.

The generalization of the ADM energy, momentum, and angular momentum for asymptotically AdS spacetime is known as the Abbott-Deser (AD) \cite{AD} formula
\begin{equation}
Q\left[\bar{\xi}\right]=\frac{1}{2\Omega_{n-2}G_{N}}\int_{\partial\bar{\Sigma}}d^{n-2}z\,\sqrt{\bar{\gamma}^{\left(\partial\bar{\Sigma}\right)}}\bar{\epsilon}_{\mu\nu}\left(\bar{\xi}_{\alpha}\bar{\nabla}_{\beta}K^{\mu\nu\alpha\beta}-K^{\mu\beta\alpha\nu}\bar{\nabla}_{\beta}\bar{\xi}_{\alpha}\right),\label{eq:Conserved_charges}
\end{equation}
where the quantities with a bar denote the background quantities, for example $\bar \xi$ is the background Killing vector and, $\bar{\epsilon}_{\mu\nu}:= \hat{n}_{(\mu} \hat{\sigma}_{\nu)}$. The "superpotential" $K^{\mu\alpha\nu\beta}$ is defined as
\begin{equation}
K^{\mu\alpha\nu\beta}:=\frac{1}{2}\left(\bar{g}^{\alpha\nu}\tilde{h}^{\mu\beta}+\bar{g}^{\mu\beta}\tilde{h}^{\alpha\nu}-\bar{g}^{\alpha\beta}\tilde{h}^{\mu\nu}-\bar{g}^{\mu\nu}\tilde{h}^{\alpha\beta}\right),\qquad\tilde{h}^{\mu\nu}:=h^{\mu\nu}-\frac{1}{2}\bar{g}^{\mu\nu}h,
\end{equation}
which in the compact notation of Kulkarni-Nomizu product \cite{Besse}, reads 
\begin{equation}
K_{\mu\alpha\nu\beta}=-\left(\bar{g} \owedge \tilde{h}\right)_{\mu\alpha\nu\beta}.
\end{equation}
Note that one can take the background metric, $g_{\mu \nu}$,  to be any Einstein spacetime, not necessarily a maximally symmetric metric (see the proof of this in \cite{Adami}). We should also note that (\ref{eq:Conserved_charges}) generalizes (\ref{adm1}) to arbitrary curvilinear coordinates in the case of flat backgrounds. This is already a good improvement over the ADM charges as one needs only one expression for all charges in "arbitrary" coordinates, instead of different expressions for each charge in Cartesian coordinates. But, of course, such a unified expression for conserved charges makes use of the time components of the vectors and tensors, not just the components on the spatial hypersurface. However, these "arbitrary" coordinates, as we shall see, must still be restricted into various classes depending on how the perturbation $h_{\mu \nu}$ decays.  So far, the moral of the story is as follows: Even though GR is a diffeomorphism-invariant theory and there is no absolute structure in it, that is, all coordinates are on an equal footing, one should not forget that the conserved charge expressions do not allow certain large diffeomorphisms that do not obey the boundary conditions. One might even say that the boundary conditions are designed to rule out these large diffeomorphisms.  This will be clear with the examples in the next section.

Before we start our detailed study of the asymptotic decay conditions of the metric, let us note that another form of conserved charges often used in the literature is \cite{Deser_Tekin-PRL,Deser_Tekin-PRD}.
\begin{equation}
Q^{\mu}\left[\bar{\xi}\right] = \frac{1}{4\Omega_{n-2}G_{n}} \int_{\partial\bar{\Sigma}} dS_{i}{\cal{F}}^{\mu i},\label{charge}
\end{equation}
where one should consider $Q^0$ as the conserved charge, and the antisymmetric integrand is
\begin{equation}
\begin{aligned}
{\cal{F}}^{\mu i}:= 
 &\bar{\xi}_{\nu} \bar{\nabla}^{0} h^{i\nu} - \bar{\xi}_{\nu} \bar{\nabla}^{i} h^{0\nu} 
+ \bar{\xi}^{0} \bar{\nabla}^{i} h - \bar{\xi}^{i} \bar{\nabla}^{0} h  + h^{0\nu} \bar{\nabla}^{i} \bar{\xi}_{\nu} 
 - h^{i\nu} \bar{\nabla}^{0} \bar{\xi}_{\nu} 
+ \bar{\xi}^{i} \bar{\nabla}_{\nu} h^{0\nu}\\
&- \bar{\xi}^{0} \bar{\nabla}_{\nu} h^{i\nu} 
+ h \bar{\nabla}^{0} \bar{\xi}^{i},
\end{aligned}\label{adt}
\end{equation} 
 which is equivalent to (\ref{eq:Conserved_charges}). The above expression can also be obtained from the symplectic structure of General Relativity (see Section IV of \cite{Nutku}). 
 
\section{The flat spacetime under large coordinate transformations}

Assuming that the topology of the spacetime allows for the ADM decomposition, the metric of the spacetime splits as 
\begin{equation}
ds^{2} 
\;=\; 
\bigl(N_{i} N^{i} - N^{2}\bigr)\,dt^{2}
\;+\; 2\,N_{i}\,dt\,dx^{i}
\;+\;\gamma_{ij}\,dx^{i}\,dx^{j},
\quad i,j \in \{1,2,3\},
\label{eq:ADMmetric}
\end{equation}
where $N = N(t,x^{i})$ is the lapse function, $N^{i} = N^{i}(t,x^{i})$ is the shift-vector; and $\gamma_{ij} = \gamma_{ij}(t,x^{j})$ is the spatial Riemannian metric which is used to lower the spatial indices. In the splitting of Einstein's equations, one realizes that the extrinsic curvature of the spatial surface defined as 
\begin{equation}
K_{ij} 
\;=\;
\frac{1}{2N}
\Bigl(
  \dot{\gamma}_{ij}
  \;-\;
  D_{i}N_{j}
  \;-\;
  D_{j}N_{i}
\Bigr),
\qquad
\dot{\gamma}_{ij} \;=\; \frac{\partial \gamma_{ij}}{\partial t},
\label{eq:ExtrinsicCurvature}
\end{equation}
plays an important role. In fact, this symmetric  together with $\gamma_{i j}$ and $\Sigma$ constitute the initial data, and it  also appears in the definitions of linear and angular momentum:
\begin{equation}
P_{i} := \frac{1}{8 \pi G_N}\, \int_{S^2_\infty} dS\, n^j \, K_{ij}, 
\hskip 1 cm 
J_{i}
\;=\;
\frac{1}{16\pi G_N}\,\varepsilon_{i j k}
\int_{S^2_\infty} dS\, n_{l}
\bigl( x^j K^{kl} \;-\; x^k K^{jl}\bigr).
\label{eq:ADMAngularMomentum}
\end{equation}
These conserved quantities together with the total energy
\begin{eqnarray}
E_{{\rm ADM}}=\frac{1}{16 \pi\,G_N}\int_{S^2_\infty} & dS_{i}\, & \left(\partial_{j}\,h^{ij}-\partial^{i}\,h_{jj}\right),
\end{eqnarray}
and the center of mass \cite{ReggeTeitelboim, BeigOMurchadha}

\begin{eqnarray}
C^{l}:=\frac{1}{16\pi\,E_{ADM}} \int_{S^2_\infty} dS\, \bigg[x^{l}\,n^j  \left(\partial_{i}\,h^{i}_{j}-\partial_{j}\,h^{i}_{i}\right)-(h^{l}_{i}\dot{n}^i-h^{i}_{i}n^l)\bigg],
\end{eqnarray}
constitute the 10 conserved quantities corresponding to the symmetries of the Poincar\'e group ISO(1,3) in 4 dimensions. Here, $n^{i}:=\frac{x^i}{r}$ and $r=\sqrt{\delta_{ij}x^ix^j}$ and $h_{ij}:=\gamma_{ij}-\delta_{ij}$. The generalization of the formulas to $n$ spacetime dimensions is straightforward.  Note that the positive energy theorem is the statement that $P^\mu= (E_{\text{ADM}}, \vec{P})$ is a time-like four vector for asymptotically flat spacetimes given that the dominant energy condition is satisfied.  

\subsection{Mass of the Flat Spacetime}

Clearly, if we consider the flat spacetime $\mathbb{R}^{1,3}$ in the Cartesian $(t, x^i)$ or the usual spherical coordinates $(t,r,\theta,\phi)$, all the conserved quantities defined above identically vanish as they should. However, starting with the metric 
\begin{equation}
ds^2_{\mathbb{R}^{1,3}} = -dt^2 + dr^2 + r^2 (d\theta^2 + \sin^2\theta\, d\phi^2), \label{vacuum}
\end{equation}
and redefining the radial coordinate $r$ as follows  \cite{DenisovSoloviev83,BrayChrusciel,Witten}
\begin{equation}
r(\rho):= \rho + c \, \rho^{1-s},\label{kemal} 
\end{equation}
with some constants $s>0$, $c\in \mathbb{R}$; and defining new the asymptotically Euclidean coordinates as \( y^i := \frac{\rho \,  x^i}{r} \), then the flat metric becomes
\begin{equation}
ds^2_{\mathbb{R}^{1,3}}= -dt^2+\delta_{ij} \, dx^i \, dx^j = -dt^2+\gamma_{ij} \, dy^i \, dy^j.
\end{equation}
The new spatial metric behaves as
\( \gamma_{ij} - \delta_{ij} = \mathcal{O}(|y|^{-s}) \), while its derivatives behave as \( \partial_k \gamma_{ij} = \mathcal{O}(|y|^{-s - 1}) \). The explicit form of $\gamma_{i j}$ can be easily computed starting from $x^{i}:=(1+c \,\rho^{-s}) \,y^{i}$, and one obtains
\begin{equation}
\gamma_{ij} = \delta_{ij} f(\rho)+y_{i}\,y_{j}\,g(\rho)
\end{equation}
where $f(\rho)$ and $g(\rho)$  are given as 
\begin{equation}
f(\rho) = ( 1 + c \rho^{-s} )^2, \hskip 1.5 cm 
g(\rho) = c\,s \, \rho^{-s-2} ( -2 \,( 1 + c \rho^{-s} ) + c\, s\, \rho^{-s} ).\label{fandg}
\end{equation}
Hence, in the spherical version of these new coordinates, the flat spacetime metric reads as
\begin{equation}
ds^2_{\mathbb{R}^{1,3}}=-dt^{2}+(f(\rho)+\rho^{2}g(\rho))\,d\rho^{2}+f(\rho)\,\rho^{2}\,d\Omega_{2}.\label{perturbed}
\end{equation}
For generic $f(\rho)$ and $g(\rho)$, this metric does not describe a flat space-time metric, but we should keep in mind that these functions are given by (\ref{fandg}). Hence, the metric (\ref{perturbed}) is diffeomorphic to the flat spacetime metric. 

Let us now calculate the ADM mass of this metric, which requires a definition of the background metric as 
\begin{equation}
 ds^2_{\text{back}}=-dt^2+d\rho^2+\rho^2d\Omega_{2},\label{vacuum2}
\end{equation}
which is obtained by setting $c=0$ which yields $g(\rho)=0$ and $f(\rho)=1$; 
and the perturbation is defined as 
\begin{equation}
h_{\mu\nu}:=g_{\mu\nu}-\bar{g}_{\mu\nu}.
\end{equation}
Inserting (\ref{perturbed}) and (\ref{vacuum2}) into the last equation, one arrives at the perturbation part of the line element.
\begin{equation}
h_{\mu\nu}dy^\mu dy^\nu = \left(-1+f(\rho)+\rho^2g(\rho)\right)\, d\rho^2+\rho^2(-1+f(\rho))\, d\Omega_{2}.
\end{equation}
It is easier to work with (\ref{adt}) as it allows the use of spherical coordinates. But, of course, one must consider the covariant derivatives with respect to the background spacetime. For example, one has
\begin{equation}
\bar{\nabla}_{\rho} \,h_{\mu \nu} = \partial_{\rho}\, h_{\mu \nu} - \bar{\Gamma}^{\lambda}_{\rho \mu}\, h_{\lambda \nu} - \bar{\Gamma}^{\lambda}_{\rho \nu}\, h_{\mu \lambda},
\end{equation}
and the scalar field $h$ is given as 
\begin{align}
h=h_{\mu\nu}\,\bar{g}^{\mu\nu}= 3f(\rho)+\rho^2 \, g(\rho) -3. 
\end{align}
The background, time-like Killing vector is $\bar{\xi}^{\mu}=(-1,0,0,0)$ which is the relevant Killing vector for the mass/energy. Inserting all these in (\ref{adt}) and keeping the radial coordinate $\rho$ finite yields a $\rho$-dependent energy:
\begin{equation}
E(\rho) = \frac{\rho \, g(\rho) - \partial_\rho f(\rho)}{2} \rho^2.\label{flatrho}
\end{equation} 
We know that we capture the total energy of spacetime only when we let $\rho\rightarrow \infty$ in this expression. But it is clear that this limit may not even exist! In fact, we have the following 3 distinct sectors of coordinate transformations of the form (\ref{kemal})
 \begin{eqnarray}
\lim_{\rho \rightarrow \infty} E= 
\begin{cases} 
\infty, &  s < \frac{1}{2}, \\
\frac{1}{8} c^2, &  s = \frac{1}{2}, \\
0, &  s > \frac{1}{2}.
\end{cases} \label{desired}
\end{eqnarray} 
Therefore, it is clear that to keep the mass of the flat spacetime to be zero, we must not allow coordinate transformations of the form (\ref{kemal}) with $s \le  1/2$. So, this decay behavior is necessary, but it turns out that it is also sufficient for the flat space to have a zero mass \cite{Bartnik}. Of course, as mentioned above, the lesson we learn from this exercise is that, even though GR is a diffeomorphism invariant theory, meaning there is no absolute structure imposed on us from the theory, and no coordinates are privileged, further fine details of the theory, such as identifying the flat Minkowski spacetime as the vacuum with zero energy/mass imposes a restriction on certain large gauge transformations. Not all diffeomorphisms keep the conserved charge expressions, as defined so far above, intact. 

This computation can easily be extended to $n$ spacetime dimensions, with the metric 
\begin{equation}
ds^{2}_{\mathbb{R}^{1,n-1}}=-dt^{2}+\left(f(\rho)+\rho^{2}\,g(\rho)\right)\,d\rho^{2}+f(\rho)\,\rho^{2}\,d\Omega_{n-2},
\end{equation}
which again splits into 3 sectors for $n>2+1$
\begin{eqnarray}
\lim_{\rho \rightarrow \infty} E= 
\begin{cases} 
\infty, &  s < \frac{n-3}{2}, \\
\frac{(n-2)(n-3)^2}{16} c^2, &  s = \frac{n-3}{2}, \\
0, &  s > \frac{n-3}{2},
\end{cases}\label{before}
\end{eqnarray}
while in $n=2+1$ dimensions, the finite energy sector disappears. One is not allowed to use coordinates that satisfy $s \le\frac{n-3}{2} $ in $n$-dimensions. 

\subsection{Angular Momentum of the Flat Spacetime}

Starting from metric (\ref{perturbed}), we make another coordinate  transformation on the azimuthal angle as follows:
\begin{equation}
    \phi = p\,(t,\rho,\psi),
\end{equation}
so the new metric (\ref{perturbed}) in the new coordinates $(t,\rho,\theta,\psi)$ becomes

\begin{eqnarray}
\begin{aligned}
ds^2_{\mathbb{R}^{1,3}}& =  \left(-1 + \rho^2 f(\rho) \sin^2\theta \, ( \partial_t p )^2 \right) \, dt^2 + 2 \rho^2 f(\rho) \sin^2\theta \, \partial_\rho p\,\partial_t p \, dt \, d\rho \\
&+ 2 \rho^2 f(\rho) \sin^2\theta \,  \partial_\psi p\, \partial_t p  \, dt \, d\psi   +  \left( f(\rho) + \rho^2 g(\rho) + \rho^2 f(\rho) \sin^2\theta \left( \partial_\rho p \right)^2 \right)  \, d\rho^2 \\
&+ 2 \rho^2 f(\rho) \sin^2\theta \,  \partial_\psi p \partial_\rho p  \, d\rho \, d\psi+ \rho^2 f(\rho) \, d\theta^2 + \rho^2 f(\rho) \sin^2\theta \left( \partial_\psi p \right)^2  \, d\psi^2.
\end{aligned}
\end{eqnarray}
Once again, as we just did coordinate transformations, this metric is diffeomorphic to the flat metric.
Carrying out the above procedure verbatim for the Killing vectors, $\xi^\mu=(-1,0,0,0)$ and $\xi^\mu=(0,0,0,1)$, and keeping $\rho$ finite, one arrives at the energy and 
\begin{equation}
 \begin{aligned}
E =&\ 
\frac{\rho^2}{8} \bigg[ 4 \, \rho \,g(\rho) 
- 2 \, \partial_\rho f(\rho) ( 1 + ( \partial_\psi \,p )^2 ) + \frac{2 f(\rho)}{3 \rho} \, \bigg(
3 - 3 \,( \partial_\psi \,p )^2
+ 3 \, \rho \,  \partial^2_{\psi} \, p \, \partial_\rho \,p \\
&\qquad + 4 \, \rho^2 \, ( \partial_\rho \, p )^2 - 3 \, \rho \, \partial_\psi \,p \, \partial_{\psi} \,\partial_{\rho} \,p\bigg) \bigg] , \label{bigE}
\end{aligned}
\end{equation}
the spin angular momentum of the new metric
\begin{equation}
 \begin{aligned}
J =&\ 
 \frac{1}{6} \rho^4 \bigg[ 
\bigg( \partial_\rho f(\rho) \, \partial_\psi p + f(\rho) \, \partial_\rho \partial_\psi p \bigg) \partial_{t} p 
- f(\rho)\, \partial_\rho \,p \, \partial_t \partial_\psi \,p
)\bigg ]. \label{bigJ}
 \end{aligned}
\end{equation}
 To be more concrete, if we choose the new azimuthal coordinate as 
\begin{equation}
\phi=p\,(t,\rho,\psi)=\psi+mt+b\rho^{-s-2},
\end{equation}
where $m$ and $b$ are dimensionful constants, then, from (\ref{bigE}) and (\ref{bigJ}), we have
\begin{align}
E&= \frac{1}{6} \rho^{-3 - 4s} \bigg[ 2\, c^2 b^2\,(2 + s)^2 + 4 \,c \,b^2\, (2 + s)^2 \rho^s + \rho^{2s} \bigg( 2\,b^2(2+s)^2+3c^2s^2\rho^4  \bigg) \bigg]
, \\
J &= 
-\frac{1}{3} m\,c \, s\, \rho^{3 - 2s} ( c + \rho^s ).
\end{align}
In this four-dimensional example, we get at spatial infinity:
\begin{equation}
\begin{array}{cc}
\begin{aligned}
\lim_{\rho \rightarrow \infty} E= 
\begin{cases} 
\infty, &  s < \frac{1}{2}, \\
\frac{c^2}{8}, &  s = \frac{1}{2}, \\
0, &  s > \frac{1}{2}.
\end{cases}
\end{aligned}
& 
\qquad
\begin{aligned}
\lim_{\rho \rightarrow \infty} J= 
\begin{cases} 
\infty, &  s < 3, \\
-mc, &  s = 3, \\
0, &  s > 3.
\end{cases}
\end{aligned}
\end{array}
\end{equation}
Therefore, it is clear that even though $s>1/2$ makes the energy of the flat-metric zero; to keep its angular momentum also zero, requires $s>3$. In $n$ dimensions, the energy part is the same as before (\ref{before}), but the angular momentum becomes:
\begin{equation}
\begin{array}{cc}
\begin{aligned}
\lim_{\rho \rightarrow \infty} J= 
\begin{cases} 
\infty, &  s < n-1, \\
-mc, &  s = n-1, \\
0, &  s > n-1.
\end{cases}
\end{aligned}
\end{array}
\end{equation}
These are some examples of large gauge transformations; one can certainly consider more complicated examples.

\section{Anti-de Sitter Spacetime under large coordinate transformations}
Let us now carry out a similar analysis for the AdS spacetime \cite{Ashtekar:1984zz} in $n$ dimensions:
\begin{equation}
ds^{2}_{\text{AdS}} = -\big(1  -\frac{2 \Lambda}{(n-2)(n-1)}r^2\big)\, dt^2 + \big(1 -\frac{2 \Lambda}{(n-2)(n-1)}r^2\big)^{-1} \, dr^2 +  r^2 \, d\Omega_{n-2},   
\end{equation}
which is a maximally symmetric space with the curvatures:
\begin{equation}
\bar{R}_{\mu\alpha\nu\beta}=\frac{2\Lambda}{\left(n-2\right)\left(n-1\right)}\,\left(\bar{g}_{\mu\nu}\,\bar{g}_{\alpha\beta}-\bar{g}_{\mu\beta}\,\bar{g}_{\alpha\nu}\right)
,\qquad
\bar{R}_{\mu\nu}=\frac{2\Lambda}{n-2}\,\bar{g}_{\mu\nu},\qquad\bar{R}=\frac{2\,n\,\Lambda}{n-2}, \label{curvatures}
\end{equation}
and $\Lambda < 0$. 
Making the transformation (\ref{kemal}), the new metric in four dimensions, becomes
\begin{equation}
ds^{2}_{\text{AdS}} = -\,\left (1 - \frac{\Lambda \, \rho^2 f(\rho)}{3}\right) \, dt^2 + \frac{f(\rho) + \rho^2 \,g(\rho)}{1-\frac{\Lambda \, \rho^2 \,f(\rho)}{3}} \, d\rho^2 + f(\rho)\, \rho^2\, d\Omega_{2},\label{adsfull}
\end{equation}
which is a maximally symmetric metric and has the same curvatures as given in (\ref{curvatures}) for $n=4$. The perturbation part of the line element is 
\begin{equation}
h_{\mu\nu}dy^\mu dy^\nu = -\frac{\Lambda\,\rho^2 \,(1-f(\rho))} {3}\, d\tau^2+\bigg(\frac{f(\rho)+\rho^2\,g(\rho)}{1-\frac{\Lambda\, f(\rho) \,\rho^2}{3}} -\frac{1}{1-\frac{\Lambda\, \rho^2}{3}}\bigg)\, d\rho^2-(1-f(\rho))\,\rho^2\, d\Omega_2.
\end{equation}
Next, we calculate the energy of this "perturbed" metric. 

\subsection{Energy/Mass of the AdS spacetime}

For the time-like Killing vector $\xi^{\mu}=(-1,0,0,0)$, from (\ref{adt})
one arrives at the energy/mass: 
\begin{align}
E= \frac{\rho^2}{2(1 -\frac{\Lambda}{3}\, \rho^2 f)}\Big[\frac{\Lambda}{3}\rho\big(1 - f\big)^2  +  \big(1- \frac{\Lambda}{3} \rho^2\big)^2 g \rho
-  (1 -\frac{\Lambda}{3} \rho^2) \big(1- \frac{\Lambda}{3}\rho^2 f\big) \partial_{\rho}f\Big],
\end{align}
where we abbreviated $f= f(\rho)$ and $g=g(\rho)$.\footnote{As expected, in $\Lambda \rightarrow 0$ limit, this expression reduces to the flat space version given as (\ref{flatrho}).} We get the following limits at the spatial infinity (note that de-Sitter spacetime does not have a spatial infinity, for that reason we have considered the AdS spacetime)
\begin{equation}
\begin{aligned}
\lim_{\rho \rightarrow \infty} E=  
\begin{cases} 
\infty, &  s < \frac{3}{2}, \\
-\frac{11}{8}\,c^2\Lambda, &  s=\frac{3}{2}, \\
0, &  s >\frac{3}{2}.
\end{cases}
\end{aligned}
\end{equation}
Therefore, to assign zero energy to the AdS geometry, we must restrict the gauge transformation of to form (\ref{kemal}) to the $s > 3/2$ sector. It is easy to generalize this to the $n$-dimensional AdS of which the results are 
\begin{equation}
\begin{aligned}
\lim_{\rho \rightarrow \infty} E=
\begin{cases} 
\infty, &   s < \frac{n-1}{2}, \\
-\frac{(n+7)}{8}\,c^2\Lambda , &  s = \frac{n-1}{2}, \\
0, &  s > \frac{n-1}{2}.
\end{cases}
\end{aligned}
\end{equation}
Therefore, we must demand the decay of the coordinates to satisfy $s> \frac{n-1}{2}$ to keep the energy of the AdS to be zero.
\subsection{Angular Momentum of the  AdS spacetime}
To study the decay constraints coming from the total angular momentum of the AdS spacetime, let us consider the following coordinate transformation
\begin{equation}
   \phi= p(t, \rho,\psi),
\end{equation}
on the metric (\ref{adsfull}) which yields
\begin{equation}
\begin{aligned}
ds^{2}_{\text{AdS}}&= \left(-1 + \frac{1}{3} \,\Lambda\, \rho^{2} f(\rho) + \rho^{2}\, f(\rho)\, \sin^{2}\theta \,\left( \partial_t p \right)^{2} \right) dt^{2} + 2\, \rho^{2} f(\rho) \sin^{2}\theta \, \partial_\rho p \, \partial_t p\, dt \, d\rho \\
&+ 2 \rho^{2} f(\rho) \sin^{2}\theta \, \partial_\psi p \, \partial_t p \, dt \, d\psi  
+ \Big(\frac{f(\rho)+\rho^{2} g(\rho)}{1 - \frac{1}{3} \Lambda \rho^{2} f(\rho)} + \rho^{2} f(\rho) \sin^{2}\theta \, (\partial_\rho p)^2  \Big) d\rho^{2} \\
&+ \rho^{2} f(\rho)  d\theta^{2}+ 2 \rho^{2}\, f(\rho)\, \sin^{2}\theta \, \partial_\psi p \, \partial_\rho p\, d\rho \, d\psi +  \rho^{2} f(\rho)\sin^{2}\theta \, (\partial_\psi p)^2 \, d\psi^{2} 
\end{aligned}
\end{equation}
Taking the coordinate transformation to be of the form,
\begin{equation}
\phi= p(t,\rho,\psi)=\psi+m\,t+b\,\rho^{-s+1/2},
\end{equation}
energy of this metric reads as
\begin{align}
E&= \bigg[54\, \Lambda \,\rho^2 + 36 \,c \,s \Lambda\,\rho^{2 - s} (3 - \Lambda \rho^2) 
        + 18\, c^2\, s \,\rho^{-2s} (3 - \Lambda \rho^2)(3 s + (2 -s)\, \Lambda \,\rho^2) \nonumber \\
        & + \Lambda\, \rho^2 \,(1 + c \,\rho^{-s})^4 \bigg(54 + b^2 (1 - 2s)^2 \rho^{1 - 2s} (3 - \Lambda\, \rho^2)^2\bigg) \nonumber\\
        & +3 \,\rho \,(1 + c \rho^{-s})^2 \bigg(-36\, \Lambda \,\rho 
            +  \,\rho^{-2s} (3 - \Lambda\, \rho^2) \big( b^2(1 - 2s)^2 (3 - \Lambda \rho^2) - 12 c s \Lambda \rho (c + \rho^s)\big)\bigg)\bigg]\nonumber \\
           & \times \frac{\rho}{108 (3 - \Lambda \,\rho^{2 - 2s} (c + \rho^s)^2)},
\end{align}
while its angular momentum reads as 
\begin{align}
J&= -\frac{1}{3}\, m\,c\, s \,\rho^{3 - 2s}\, (c +\rho^{-s}).
\end{align}
In $n$ spacetime dimensions, we obtain the following limits
\begin{equation}
\begin{array}{cc}
\begin{aligned}
\lim_{\rho \rightarrow \infty} E= 
\begin{cases} 
\infty, &   s < \frac{n+2}{2}, \\
\frac{(n+1)^2}{2^{\,n/2}(n-1)^3}b^2\Lambda^2, & \text{if } n \text{ is even}, \quad s = \frac{n+2}{2},\\
\frac{n+1}{2\left[(n-2)^3+\frac{5}{2}(n-3)\right]}b^2\Lambda^2, & \text{if } n \text{ is odd}, \quad  s = \frac{n+2}{2},\\
0, &  s > \frac{n+2}{2}.
\end{cases}
\end{aligned}
& 
\qquad
\begin{aligned}
\lim_{\rho \rightarrow \infty} J= 
\begin{cases} 
\infty, &  s < n-1, \\
-m\,c, &  s = n-1, \\
0, &  s > n-1.
\end{cases}
\end{aligned}
\end{array}
\end{equation}
In four dimensions, one must have $s>4$ to assign a zero mass and angular momentum to the AdS metric.

\subsection{Kerr-Schild form of the AdS spacetime}

Kerr-Schild coordinates are very suitable for many black hole solutions \cite{Gurses:1975vu,Dereli:1986cm}. Let us study here the Kerr-Schild form of the AdS spacetime.

\begin{equation}
    ds_{\text{AdS}}^2=\eta_{\mu\nu} dx^{\mu} dx^{\nu} +(1-a(r))(k_{\mu} dx^{\mu})^{2}
\end{equation}
where $a(r)=1-\Lambda\, r^2 /3$, and $k^\mu$ is a null vector with respect to both $\eta_{\mu \nu}$ and the full metric $g_{\mu \nu}$. Doing the coordinate transformation (\ref{kemal}), one has 
\begin{equation}
  \begin{aligned}  
  ds_{\text{AdS}}^{2}=&-dt{^2}+[f(\rho)+\rho^2 \,g(\rho)]\,d\rho^2 + f(\rho)\,\rho^{2}\,d\Omega_{2}+\,(1-a(\rho))\,[\,dt-k(\rho)\,d\rho\,]^{2},
\end{aligned}\label{adsnew}
 \end{equation}   
where $k(\rho)$ is defined as
\begin{equation}
   k(\rho)=1+(1-s)\,c\,\rho^{-s}. 
\end{equation}
Note that (\ref{adsnew}) is the AdS spacetime with these "goofy coordinates". 
We take the background metric $\bar{g}_{\mu\nu}$ to be 
\begin{equation}
ds_{\text{back}}^{2}=-dt^2+d\rho^2+\rho^2\,d\Omega_{2}+\frac{\rho^2\Lambda}{3} (\,dt-d\rho)^2,
\end{equation}
and the deviation to be 
\begin{equation}
\begin{aligned}
    h_{\mu\nu}dy^\mu dy^\nu &=  \bigg(-\frac{\Lambda \,\rho^2 (1 - f)}{3}\bigg)  \,dt^2 
    +   \bigg(\frac{2\,\Lambda\, \rho^2(1 - f\,k)}{3}\bigg)  dt \,d\rho  \\
    &+ \bigg( f(1+\frac{\Lambda \,\rho^2 \,k^2}{3}) 
     + \rho^2\, g-\frac{3+\rho^2\Lambda}{3} \bigg) \,d\rho^2 +  \left( -1 + f\,\right)\rho^2  d\Omega_{2}.
\end{aligned}
\end{equation}
Computing the energy using (\ref{adt}) for this metric, one arrives at 
\begin{equation}
\begin{aligned}
E=&\frac{\rho^2}{54} (-3 + \Lambda \,\rho^2)  \Big[ -3\,\Lambda\,\rho 
    + 3\,\rho (-3 + \Lambda\, \rho^2) g(\rho) + \Lambda\, \rho\, f(\rho) \,\Big( 6 + \,\Lambda \,\rho^2 
    - 2 \Lambda\, \rho^2\, k(\rho) \\
    &+ (-3 + \Lambda\, \rho^2) \,k(\rho)^2 \Big) 
    + 9 \,\partial_{\rho}f(\rho) \Big]. \label{uzun}
\end{aligned}
\end{equation}
As in the previous section, $\Lambda \rightarrow 0$ limit of the expression yields the flat space case (\ref{flatrho}).
At spatial infinity from (\ref{uzun}), we get
\begin{eqnarray}
\lim_{\rho \rightarrow \infty}E= 
\begin{cases} 
-\infty, &  s < \frac{7}{2}, \\
\frac{25}{216} c^2 \Lambda^3 , &  s = \frac{7}{2}, \\
0, &  s > \frac{7}{2}.
\end{cases}
\end{eqnarray}
This can be generalized to $n$ dimensions whose details we do not depict here. One obtains
\begin{equation}
\begin{array}{cc}
\begin{aligned}
\lim_{\rho \rightarrow \infty}E= 
\begin{cases} 
\infty, &   s < \frac{n+3}{2}, \\
\frac{(n+1)^2}{2^{n-1}(n-1)^3}c^2\,\Lambda^3, & \text{if } n \text{ is even}, \quad s = \frac{n+3}{2},\\
\frac{n+1}{(n-2)[2(n-2)^3+5(n-3)]}c^2\,\Lambda^3, & \text{if } n \text{ is odd}, \quad  s = \frac{n+3}{2},\\
0, &  s > \frac{n+3}{2}.
\end{cases}
\end{aligned}
\end{array}
\end{equation}
Finally, to see the behavior of the angular momentum, let us consider the new coordinates, in addition to the ones defined in (\ref{kemal}), 
\begin{equation}
 \phi=p(\tau,\rho,\psi), \quad t=y(\tau,\rho),   
\end{equation}
The angular momentum becomes
\begin{equation}
J = \frac{\rho^4}{6} \bigg( \partial_\rho f(\rho) \, \partial_\psi  p
+ f(\rho)\, \partial_\rho \partial_\psi\, p \partial_\tau p
- f(\rho) \partial_\rho p \,\partial_\tau \partial_\psi p\bigg).
\end{equation}
Choosing the functions as 
\begin{equation}
 p(\tau,\rho,\psi)=b\tau+\psi+m\rho^{-s}, \quad \quad \quad y(\tau,\rho)=l\,\rho,
 \end{equation}
we get
\begin{equation}
J=-\frac{1}{3}\, c\,b\, s \,\rho^{3 - 2 s} \,(c + \rho^s).
\end{equation}
In $n$ dimensions, one has 
\begin{equation}
 \begin{aligned}
\lim_{\rho \rightarrow \infty} J= 
\begin{cases} 
\infty, &  s < n-1, \\
-c\,b, &  s = n-1, \\
0, &  s > n-1.
\end{cases}
\end{aligned}
\end{equation}
Once again, keeping the energy and the angular momentum of the AdS vacuum to be zero requires these decay conditions on the coordinates.

All of the above computations teach us that we must be careful in using the well-established conserved charge definitions in gravity theories. By changing the coordinates that do not decay properly, one can change the charges assigned to the vacuum of the theory, even though the coordinate changes are diffeomorphisms, which are not expected to affect the physical results. Here, we studied the flat and AdS spacetimes as examples, but a similar computation can be done for non-vacuum solutions, such as black holes. Of course, the crux of the problem is that the conserved charges written above have integrands that contain the first derivatives of the metric or the first derivatives of the deviations of the metric from a background metric. One might try to remedy this problem by finding a formula where the integrand does not contain the first derivative but contains the second derivative of the metric; i.e., the Riemann tensor. Next, we describe this in the context of asymptotically AdS spacetimes, both as solutions to cosmological Einstein's theory and higher derivative theories. Unfortunately, currently, we do not know how to carry out a similar calculation for asymptotically flat backgrounds. The asymptotically AdS case was found in \cite{altas1,altas2}. Of course, to obtain the asymptotically flat case, one can take the limit $\Lambda \rightarrow 0$ at the end of the computations, but still, one would like to have a compact formula of conserved charges in asymptotically flat spacetimes in terms of the curvature. This is still an open problem for us. 

\section{Resolving the large gauge transformation problem in AdS}

The following construction pertaining to the definition of conserved charges was done in detail in (\cite{altas1,altas2}), therefore, we just want to recapitulate the salient features. The idea starts from the following simple, apparently unrelated, question: The Ricci tensor is the trace of the Riemann tensor; is there a rank $(1,3)$ tensor of which the trace is the Einstein tensor?  The answer is affirmative for dimensions $n>3$ and is unique up to a "constant" part that can be fixed as desired. Here is the explicit form of that tensor, which we called the "${{\cal {P}}}$-tensor for the lack of a better name:
\begin{equation}
\text{\ensuremath{{\cal {P}}}}^{\nu}\thinspace_{\mu\beta\sigma}:=R^{\nu}\thinspace_{\mu\beta\sigma}+\delta_{\sigma}^{\nu}\text{\ensuremath{{\cal {G}}}}_{\beta\mu}-\delta_{\beta}^{\nu}\text{\ensuremath{{\cal {G}}}}_{\sigma\mu}+\text{\ensuremath{{\cal {G}}}}_{\sigma}^{\nu}g_{\beta\mu}-\text{\ensuremath{{\cal {G}}}}_{\beta}^{\nu}g_{\sigma\mu}+\left(\frac{R}{2}-\frac{\Lambda\left(n+1\right)}{n-1}\right)\left(\delta_{\sigma}^{\nu}g_{\beta\mu}-\delta_{\beta}^{\nu}g_{\sigma\mu}\right),
\end{equation}
where  ${\mathcal{G}}_{\mu \nu}:= R_{\mu \nu} -\frac{1}{2} g_{\mu \nu} R + \Lambda g_{\mu \nu}$ is the cosmological Einstein tensor; and we added the "constant" part which is the explicitly $\Lambda$-dependent part to set ${{\cal {P}}}^{\nu}\thinspace_{\mu\beta\sigma}=0$ for the maximally symmetric background. Contracting the tensor once, one gets the cosmological tensor
\begin{equation}
    {{\cal {P}}}^{\sigma}\,_{\mu\sigma\nu}= -(n-3) {\mathcal{G}}_{\mu \nu}, 
\end{equation}
as was the original motivation of the construction of this tensor. Note that in $n=2+1$, the Ricci and Riemann tensors carry the same amount of information, and the ${{\cal {P}}}$ -tensor is identically zero. 

The most important property of the ${\cal {P}}$-tensor is that it is divergence-free everywhere in spacetime for any theory:
\begin{equation}
\nabla_\nu\text{\ensuremath{{\cal {P}}}}^{\nu}\thinspace_{\mu\beta\sigma}=0, \hskip 1 cm \text{for all metrics}.\label{div}
\end{equation}
It satisfies the algebraic Bianchi identity (i.e. ${\cal {P}}_{(\mu \nu \sigma) \rho }=0$),   just like the Riemann tensor, but it does not satisfy the differential Bianchi identity (i.e. $\nabla_{(\lambda} {\cal {P}}_{\mu \nu ) \sigma \rho }\ne 0$),  instead it is divergence-free as noted in (\ref{div}). It is important to note the following: in Einstein's theory, the Riemann tensor is divergence-free only in a vacuum outside the matter sources, but the divergence-free property of the ${{\cal {P}}}$ is a Bianchi-Identity valid everywhere and for any theory.  So compare the below equation with (\ref{div})
\begin{equation}
\nabla_\nu R^{\nu}\thinspace_{\mu\beta\sigma}=0, \hskip 1 cm \text{for  Einstein metrics}.\label{div2}
\end{equation}
The divergence property of the ${{\cal {P}}}$ (\ref{div}) is very useful for constructing the conserved quantities as we show below.

Consider now a generic gravity theory defined by the field equations coming from a diffeomorphism-invariant action so that its covariant derivative is identically zero:
\begin{equation}
 \mathcal{E}^{\mu \nu} = \kappa T_{\mu \nu}, \hskip 1 cm  \nabla_\mu  \mathcal{E}^{\mu \nu} =0, \label{EOM}
\end{equation}
and let $\bar{g}_{\mu\nu}$ be the background solution (for the $T_{\mu \nu}=0$ case) with at least one time-like Killing vector $\bar{\xi}$. [Without a time-like Killing vector, which exists at least outside a finite-radius ball in spacetime, we do not know what it means to have a conserved charge.]
Then, we define a perturbed metric around the background metric as 
\begin{equation}
    g_{\mu\nu}=\bar{g}_{\mu\nu}+\,h_{\mu\nu},
\end{equation}
which can be used to split the field equations in such a way that only the linear term is kept on the left-hand side, while all the other terms are carried to the right-hand side as contributions to the source part.  So, non-linearity of gravity is not oversimplified by this linearization procedure: the non-linear terms, $\mathcal{O}(h_{\mu \nu}^{(1+k)})$ with $k>1$, all contribute to the source as the self-gravitation of the gravitational field. This procedure gives us a covariantly conserved background tensor $\bar{\nabla}_\mu (\mathcal{E}^{\mu\nu})^{(1)}=0$ as the linearization of ${\nabla}_\mu (\mathcal{E}^{\mu\nu})=0$. To get a current, that is ordinarily conserved, out of this, we need the Killing vector:
\begin{equation}
 \mathcal{J}^{\mu}:=\sqrt{-\bar{g}}\;\bar{\xi}_{\nu}\,(\mathcal{E}^{\mu\nu})^{(1)}, \hskip 1 cm \partial_\mu  \mathcal{J}^{\mu}=0.
\end{equation}
Here, partial conservation is important, as opposed to covariant conservation, since we would like to use Stokes' theorem. Hence integrating over a spacelike hypersurface $\bar{\Sigma}$, with a time-like unit normal $\hat{n}_\mu$, yields the conserved charge: 
\begin{equation}
    Q(\bar{\xi}):=\int_{\bar{\Sigma}}d^{n-1}y\,\sqrt{\bar{\gamma}}\;\hat{n}_{\mu}\,\bar{\xi}_{\nu}\,(\mathcal{E}^{\mu\nu})^{(1)},\label{conserved}
\end{equation}
where $\gamma$ is the Riemannian metric on $\bar{\Sigma}$ pulled back from the spacetime using the embedding map 
$\Phi:\bar{\Sigma} \rightarrow {\mathcal{M}}$. We will now consider Einstein's theory for which $\mathcal{E}_{\mu \nu} ={\mathcal{G}}_{\mu \nu}$. Then, one can show the following relation \cite{altas1,altas2}:
\begin{equation}
\bar{\text{\ensuremath{\xi}}}_{\lambda}(\text{\ensuremath{{\cal {G}}}}^{\lambda\mu})^{\left(1\right)}=\frac{(n-1)(n-2)}{4\Lambda\left(n-3\right)}\bar{\nabla_{\nu}}\biggl((\text{\ensuremath{{\cal {P}}}}^{\nu\mu\beta\sigma})^{\left(1\right)}\bar{\nabla}_{\beta}\bar{\xi}_{\sigma}\biggr).\label{eq:finallinearequation}
\end{equation}
This is an important identity: it provides us with the following conserved current 
\begin{equation}
  \mathcal{J}^{\mu}=   \sqrt{-\bar{g}}\bar{\text{\ensuremath{\xi}}}_{\lambda}(\text{\ensuremath{{\cal {G}}}}^{\lambda\mu})^{\left(1\right)}=\frac{(n-1)(n-2)}{4\Lambda\left(n-3\right)}\partial_{\nu}\biggl(\sqrt{-\bar{g}}(\text{\ensuremath{{\cal {P}}}}^{\nu\mu\beta\sigma})^{\left(1\right)}\bar{\nabla}_{\beta}\bar{\xi}_{\sigma}\biggr),
\end{equation}
in terms of $(\text{\ensuremath{{\cal {P}}}}^{\nu\mu\beta\sigma})^{\left(1\right)}$ which reads explicitly as 
\begin{eqnarray}
({\cal {P}}^{\nu\mu\beta\sigma})^{\left(1\right)}= &  & (R^{\nu\mu\beta\sigma})^{1}+2(\text{\ensuremath{{\cal {G}}}}^{\mu[\beta})^{(1)}\overline{g}^{\sigma]\nu}+2(\text{\ensuremath{{\cal {G}}}}^{\nu[\sigma})^{(1)}\overline{g}^{\beta]\mu}+(R)^{\left(1\right)}\overline{g}^{\mu[\beta}\overline{g}^{\sigma]\nu}\nonumber \\
 &  & +\frac{4\Lambda}{(n-1)(n-2)}(h^{\mu[\sigma}\overline{g}^{\beta]\nu}+\overline{g}^{\mu[\sigma}{h}^{\beta]\nu}),\label{ktensorlinear}
\end{eqnarray}
where we have used the anti-symmetrization notation with a factor of 1/2.  We can use Stokes' theorem one more time to 
reduce our integral over the spatial "volume" $\bar{\Sigma}$ to its boundary $\partial \bar{\Sigma}$ with a spacelike unit normal $\hat{\sigma}_\mu$ to arrive at the conserved charge
\begin{equation}
\phantom{\frac{\frac{\xi}{\xi}}{\frac{\xi}{\xi}}}Q\left(\bar{\xi}\right)=\frac{(n-1)(n-2)}{8(n-3)\Lambda G\Omega_{n-2}}\int_{\partial\bar{\Sigma}}d^{n-2}x\,\sqrt{\bar{\gamma}}\,\bar{\epsilon}_{\mu\nu}\left(R^{\nu\mu}\thinspace_{\beta\sigma}\right)^{\left(1\right)}\bar{\nabla}^\beta \bar{\xi},^\sigma\label{newcharge}
\end{equation}
wherein the ${\mathcal{P}}$-tensor reduces to the Riemann tensor (actually to the Weyl tensor) for Einstein spaces at spatial infinity. Details of the construction can be found in (\cite{Bayram}). Under gauge transformations $\delta_\zeta h_{\mu \nu} = \mathcal{L}_\zeta g_{\mu \nu}= \nabla_\mu \zeta_\nu + \nabla_\nu \zeta_\mu $, it is easy to see that 
\begin{equation}
\delta_\zeta\left(R^{\nu\mu}\thinspace_{\beta\sigma}\right)^{\left(1\right)}= \mathcal{L}_\zeta \bar{R}^{\nu\mu}\thinspace_{\beta\sigma}=0,
\end{equation}
where the last equality follows for maximally symmetric backgrounds. The main difference between (\ref{adt}) and
(\ref{newcharge}) is that the latter has an explicitly gauge-invariant integrand, while the former changes under gauge transformations. Under gauge transformations, they are related up to a boundary term as was studied in (\cite{altas1,altas2}). If the gauge transformations are not large gauge transformations as discussed above, these two expressions give the same results. However, if one considers large gauge transformations, they yield different results. The advantage of  (\ref{newcharge}) is that even large gauge transformations do not change the result. For example, for the AdS metrics we discussed above, in any coordinate system (\ref{newcharge}) yields zero charges as desired for the vacuum of the theory. 

\section{Conclusions and Further Discussions}

We have revisited here an often overlooked aspect of the conserved quantities, the fact that their usual formulation requires some stringent decay conditions of the dynamical fields, in Einstein gravity, both with and without a cosmological constant, for both asymptotically flat and AdS spacetimes. A good conserved charge definition requires at least one timelike Killing vector, a background spacetime of which all charges are assumed to be zero, and a solution that asymptotically approaches the background geometry. Of course, in principle, GR is a theory without an absolute structure, and no coordinates are privileged; therefore, the charge definitions should not depend on the coordinates used. For example, and this is probably the most crucial point here, the assignment of zero conserved charges to the background should be a coordinate-invariant statement. This quite natural expectation turns out to be not realized: we know that large gauge transformations can change the zero charge of the background to any desired value, as can be seen from (\ref{desired}). We must recall that all the metrics in that equation are exactly flat in the sense of the Riemann curvature tensor being identically zero. Yet, as far as the ADM mass is concerned, there are 3 distinct super-selection sectors of flat spacetime: the infinite-mass, finite-mass, and zero-mass spacetimes. The finite mass is a one-parameter family as the number $c$ can take any real value. One way to read all of this is to restrict the possible large-gauge transformations to the sector that leaves the mass of the vacuum to be zero. This was already noted by Bartnik \cite{Bartnik} who showed that
the ADM mass is a geometric invariant of a 3-manifold that is asymptotically flat and the metric decays as $h_{ ij } = \delta_{ij} +{\mathcal{O}}(r^{-1/2+ \epsilon})$ with $\epsilon >0$. Observe that this decay is much weaker than, say, the decay of the Schwarzschild metric in isotropic coordinates. So, from this point of view, coordinates are restricted in GR: one must use coordinates that give the ADM mass of the flat spacetime to be zero. From another point of view, one might try to answer the following question: Can one write down conserved charge formulas that involve gauge-invariant integrands? This was answered affirmatively for AdS backgrounds in \cite{altas1,altas2}, and we studied this above. 

The problem of assigning zero conserved charges to the vacuum of the theory we discussed above acquires another, rather non-trivial complication in higher curvature theories of gravity. For the sake of brevity, let us consider a purely geometric theory without additional fields, such as non-minimally coupled scalar fields.  Such a theory can be schematically described by the action
\begin{equation}
S=\int d^{n}x \sqrt{-g}\,\Bigg ( \frac{1}{\kappa}\left(R-2\Lambda_{0}\right)+\sum_{p=2}^{\infty}a_{p}\Big(\text{Riem, Ric, R, }\nabla\text{Riem, }\dots\Big)^{p}\Bigg),\label{generic_higher}\end{equation}
whose origin we are not interested in, but it can represent a low-energy theory of a more fundamental theory, such as string theory, for which we would know how to compute coefficients $a_p$ in principle in perturbation theory. One might employ the Killing charge construction here and quite easily see that all the terms in the action generically will contribute to the conserved charges. In fact, as a concrete example, let us consider quadratic gravity \cite{Deser_Tekin-PRL,Deser_Tekin-PRD} 
\noindent 
\begin{equation}
S=\int d^{n}x\,\sqrt{-g}\left[\frac{1}{\kappa}\left(R-2\Lambda_{0}\right)+\alpha R^{2}+\beta R^{\mu\nu}R_{\mu\nu}+\gamma\left(R^{\mu\nu\rho\sigma}R_{\mu\nu\rho\sigma}-4R^{\mu\nu}R_{\mu\nu}+R^{2}\right)\right].\label{eq:Quadratic_action}
\end{equation}
The theory has generically 2 different maximally symmetric vacua given by the quadratic equation
\begin{equation}
\frac{\Lambda-\Lambda_{0}}{2\kappa}+\left[\left(n\alpha+\beta\right)\frac{\left(n-4\right)}{\left(n-2\right)^{2}}+\gamma\frac{\left(n-3\right)\left(n-4\right)}{\left(n-1\right)\left(n-2\right)}\right]\Lambda^{2}=0.\label{quadratic}
\end{equation}
Let us now assume that we consider one of these two vacua as the background spacetime (that is, globally AdS) with zero conserved charges. Then, for any other solution that is asymptotically AdS, the conserved charges are given as \cite{Deser_Tekin-PRL,Deser_Tekin-PRD} 
\begin{align}
Q_{\text{quadratic}}[\bar{\xi}]= & \left(\frac{1}{\kappa}+\frac{4\Lambda n}{n-2}\alpha+\frac{4\Lambda}{n-2}\beta+\frac{4\Lambda\left(n-3\right)\left(n-4\right)}{\left(n-1\right)\left(n-2\right)}\gamma\right)Q_{\text{Einstein}}[\bar{\xi}],\label{eq:Quad_charge}
\end{align}
so the conserved charges in this theory for any solution, that is asymptotically AdS, are given by the front factor multiplied by the charge of that solution computed in Einstein's theory. One might realize that the parameter space could be in such a way that the front factor vanishes identically, namely 
\begin{equation}
\frac{1}{\kappa}+\frac{4\Lambda n}{n-2}\alpha+\frac{4\Lambda}{n-2}\beta+\frac{4\Lambda\left(n-3\right)\left(n-4\right)}{\left(n-1\right)\left(n-2\right)}\gamma=0,
\end{equation}
which leads to the following conundrum: the conserved charges of this quadratic theory for any asymptotically AdS spacetime are exactly zero,  just like the globally AdS spacetime. In fact, such a theory was studied and dubbed "Critical Gravity" in \cite{pope1,pope2}. Non-vacuum solutions having all the charges of the vacuum are rather hard to understand. Such a phenomenon was noted in a four-dimensional purely quadratic gravity for asymptotically flat spacetimes in \cite{Strominger} and was interpreted as confinement of energy in gravity. In \cite{Deser_Tekin-PRD}, a section is devoted to zero-energy models for asymptotically AdS spacetimes. Particle spectrum of these theories at the critical point suggests that a new branch of solutions arises (which are usually logarithmic in the usual coordinates) that have lower energy than the vacuum \cite{bt}. Therefore, it is highly likely that these theories in which the vacuum and non-vacuum solutions are degenerate as far as their charges are concerned do not have a stable vacuum.

Finally, let us briefly note a topic that has rather interesting ramifications but requires a much longer discussion: That is, the relation between the large gauge transformations discussed here and the supertranslations and their connections to soft charges. Diffeomorphisms that do not die off at infinity (such as the large diffeomorphisms discussed here ) change the physical state and must be treated carefully, as we have seen. In general, we can divide them into 
\begin{itemize}
    \item \textbf{BMS \cite{BMS1,BMS2} supertranslations:} large diffeomorphisms at null infinity. These are angle-dependent translations of the null coordinate.
    \item \textbf{Asymptotic symmetries at spatial infinity:} governed by parity conditions \cite{Regge}.
\end{itemize}
At null infinity, the BMS supertranslations survive and generate physically distinct vacua. At spatial infinity, supertranslations are gauge redundancies if parity conditions are imposed.
Relaxing parity at $i^0$ connects to BMS-like symmetries but challenges the ADM construction as some charges diverge.
Supertranslations and soft theorems highlight infrared subtleties in gravitational theories. They indicate a rich vacuum structure and are also related to gravitational memory effects. ADM charges, well defined under stricter falloffs, obscure this structure unless parity conditions are relaxed. Reconciling spatial and null infinity remains an active area of research. See \cite{Geiller} for further discussions.

\section{Acknowledgments}
L. Ogurol is supported by TUBITAK 2211-A.

\end{document}